\documentclass[a4paper,10pt]{article}

\usepackage{amsmath,amsgen,latexsym}
\usepackage{amstext,amssymb,amsfonts,latexsym}
\usepackage{theorem}
\usepackage{pifont}
\usepackage{microtype}
\usepackage{graphicx}
\usepackage[hyphens]{url}

 \newcommand{\bs}{\bigskip}
 \newcommand{\ms}{\medskip}
 \newcommand{\n}{\noindent}
 \newcommand{\s}{\smallskip}
 \newcommand{\hs}[1]{\hspace*{ #1 mm}}
 \newcommand{\vs}[1]{\vspace*{ #1 mm}}

 \newcommand{\setempty}{\varnothing}
 
 \newcommand{\nat}{\mathbb{N}}
 
 \newcommand{\integer}{\mathbb{Z}}

 \newcommand{\co}{\mathrm{co}\mbox{-}}

 \newcommand{\PP}{{\cal P}}

 \newcommand{\dl}{\mathrm{L}}
 \newcommand{\nl}{\mathrm{NL}}
 \newcommand{\p}{\mathrm{P}}
 \newcommand{\np}{\mathrm{NP}}

\theoremstyle{plain}
\theoremheaderfont{\bfseries}
 \newtheorem{theorem}{Theorem}[section]
 \newtheorem{lemma}[theorem]{Lemma}
 \newtheorem{proposition}[theorem]{{\bf Proposition}}
 \newtheorem{corollary}[theorem]{Corollary}

 {\theorembodyfont{\rmfamily}
  \newtheorem{definition}[theorem]{Definition}}
 {\theorembodyfont{\rmfamily} }
 {\theorembodyfont{\rmfamily} }

 \newenvironment{yproof}{\par \noindent
            {\bf Proof. \hs{2}}}{\hfill$\Box$ \vspace*{3mm}}

 \newcommand{\pair}[1]{\langle #1 \rangle}

\newcommand{\ignore}[1]{}

\newcommand{\cent}{|\!\! \mathrm{c}}
\newcommand{\dollar}{\$}

 \newcommand{\psublin}{\mathrm{PsubLIN}}
 \newcommand{\dstcon}{\mathrm{DSTCON}}

 \newcommand{\sLreduces}{\leq^{\mathrm{sL}}_{m}}
 
 \newcommand{\sLTreduces}{\leq^{\mathrm{sL}}_{T}}

\begin{document}

\pagestyle{plain}
\setcounter{page}{1}

\begin{center}
{\Large {\bf Parameterized-NL Completeness of Combinatorial Problems by Short Logarithmic-Space Reductions and Immediate Consequences of the Linear Space Hypothesis}}\footnote{A conference version will appear in the Proceedings of the Future Technologies Conference (FTC 2022), October 2022.}
\bs\\

{\sc Tomoyuki Yamakami}\footnote{Current Affiliation: Faculty of Engineering, University of Fukui, 3-9-1 Bunkyo, Fukui 910-8507,  Japan} 
\end{center}



\begin{abstract}
The concept of space-bounded computability has become significantly important in handling vast data sets on memory-limited computing devices. To replenish the existing short list of NL-complete problems whose instance sizes are dictated by log-space size parameters, we propose new additions obtained directly from natural parameterizations of three
typical NP-complete problems---the vertex cover problem, the exact cover by 3-sets problem, and the 3-dimensional matching problem.
With appropriate restrictions imposed on their instances, the proposed  decision problems parameterized by appropriate size parameters are proven to be equivalent in computational complexity to either the parameterized $3$-bounded 2CNF Boolean formula satisfiability problem or the parameterized degree-$3$ directed $s$-$t$ connectivity problem by ``short'' logarithmic-space reductions.
Under the assumption of the linear space hypothesis, furthermore, none of the proposed problems can be solved in polynomial time if the memory usage is limited to sub-linear space.

\ms
\n{\bf Key words.}
{parameterized decision problem, linear space hypothesis, NL-complete problem, sub-linear space, 2SAT, vertex cover, exact cover, perfect matching}
\end{abstract}


\sloppy
\section{Background and New Challenges}\label{sec:introduction}

\subsection{Combinatorial Problems and NL-Completeness}\label{sec:graph-related-problem}

Given a combinatorial problem, it is desirable, for practical reason, to seek for good algorithms that consume \emph{fewer} computational resources in order to solve the problem, and therefore it is of great importance for us to identify the smallest amount of computational resources required to execute such algorithms.
Of various resources, we are focused in this exposition on the smallest ``memory space'' used by an algorithm that runs within certain reasonable ``time span.''
The study on the minimal memory space has attracted significant attention in real-life circumstances at which we need to manage vast data sets  for most network users who operate memory-limited computing devices. It is therefore useful in general to concentrate on the study of space-bounded computability within reasonable execution time. In the past literature, special attention has been paid to polynomial-time algorithms using logarithmic memory space and two corresponding space-bounded complexity classes: $\dl$ (deterministic logarithmic space) and $\nl$ (nondeterministic logarithmic space).

In association with $\dl$ and $\nl$, various combinatorial problems have been discussed by, for instance,  Jones, Lien, and Laaser \cite{JLL76},  Cook and McKenzie \cite{CM87}, \`{A}lvarez and Greenlaw \cite{AG00}, and Jenner \cite{Jen95}.
Many graph properties, in particular, can be algorithmically checked using small memory space.
Using only logarithmic space\footnote{Those problems were proven to be in $\co\mathrm{SL}$ by Reif \cite{Rei84} and $\mathrm{SL}$-complete by \`{A}lvarez and Greenlaw \cite{AG00}, where $\mathrm{SL}$ is the symmetric version of $\nl$. Since $\mathrm{SL}=\co\mathrm{SL}=\dl$ by Reingold \cite{Rei08}, nonetheless, all the problems are in $\dl$.} (cf. \cite{AG00,Rei84}),  for instance, we can easily solve the problems of determining whether or not a given graph is a bipartite graph, a computability graph, a chordal graph, an interval graph, and a split graph. On the contrary, the \emph{directed $s$-$t$ connectivity problem} (DSTCON) and the \emph{2CNF Boolean formula satisfiability problem} (2SAT) are known  to be $\nl$-complete \cite{JLL76} (together with the result of \cite{Imm88,Sze88}) and seem to be unsolvable using logarithmic space. To understand the nature of $\nl$ better, it is greatly beneficial to study more interesting problems that fall into the category of $\nl$-complete problems.

\subsection{Parameterization of Problems and the Linear Space Hypothesis}\label{sec:linear-space-hypothesis}

Given a target combinatorial problem, it is quite useful from a practical viewpoint to ``parameterize'' the problem by introducing an adequate ``size parameter'' as a unit basis of measuring the total amount of computational resources, such as runtime and memory space needed to solve the problem. As quick examples of size parameters, given a graph instance $G$, $m_{ver}(G)$ and $m_{edg}(G)$ respectively denote the total number of the vertices of $G$ and the total number of the edges in $G$.
For a CNF Boolean formula $\phi$, in addition, $m_{vbl}(\phi)$ and $m_{cls}(\phi)$ respectively express the total number of distinct variables in $\phi$ and the total number of clauses in $\phi$.
Decision problems coupled with appropriately chosen size parameters are generally referred to as \emph{parameterized decision problems}.
In this exposition, we are particularly interested in $\nl$ problems $L$ parameterized by \emph{log-space (computable) size parameters} $m(x)$ of input $x$. Those parameterized decision problems are succinctly denoted $(L,m)$, in particular, to emphasize the size parameter $m(x)$. Their precise definition will be given in Section \ref{sec:parameterized}.

Among all parameterized decision problems with log-space size parameters $m$, we are focused on combinatorial problems $L$ that can be solvable by appropriately designed polynomial-time algorithms using only \emph{sub-linear space}, where the informal term ``sub-linear'' means $O(m(x)^{1-\varepsilon}polylog(|x|))$ for an appropriately chosen constant  $\varepsilon\in(0,1]$ independent of $x$. All such parameterized decision problems $(L,m)$ form the complexity class $\psublin$ \cite{Yam17a} (see Section \ref{sec:parameterized} for more details). It is natural to ask if all $\nl$ problems parameterized by log-space size parameters (or briefly, \emph{parameterized-$\nl$ problems}) are solvable in polynomial time using sub-linear space.
To tackle this important question, we zero in on the most difficult (or ``complete'') parameterized-$\nl$ problems.  As a typical example, let us consider the \emph{$3$-bounded 2SAT}, denoted $\mathrm{2SAT}_3$, for which every variable in a 2CNF Boolean formula $\phi$ appears at most 3 times in the form of literals,
parameterized by $m_{vbl}(\phi)$ (succinctly,  $(\mathrm{2SAT}_3,m_{vbl})$). It was proven in \cite{Yam18} that $(\mathrm{2SAT}_3,m_{vbl})$ is complete for the class of all parameterized-$\nl$ problems.

Lately, a practical working hypothesis, known as the \emph{linear space hypothesis} \cite{Yam17a}, was proposed in connection to the computational hardness  of parameterized-$\nl$  problems. This  linear space hypothesis (LSH) asserts that $(\mathrm{2SAT}_3,m_{vbl})$ cannot be solved by any polynomial-time algorithm using sub-linear space.
From the $\nl$-completeness of $\mathrm{2SAT}_3$, LSH immediately derives long-awaited complexity-class separations, including $\dl\neq\nl$ and $\mathrm{LOGDCFL}\neq \mathrm{LOGCFL}$, where $\mathrm{LOGDCFL}$ and $\mathrm{LOGCFL}$ are respectively the log-space many-one closure of $\mathrm{DCFL}$ (deterministic context-free language class) and $\mathrm{CFL}$ (context-free language class)  \cite{Yam17a}. Moreover,  under the assumption of LSH, it follows that 2-way nondeterministic finite automata are simulated by ``narrow'' alternating finite automata \cite{Yam19b}.

Notice that the completeness notion requires ``reductions'' between two problems. The standard $\nl$-completeness notion uses logarithmic-space (or log-space) reductions.
Those standard reductions, however, seem to be too powerful to use in many real-life circumstances. Furthermore, $\psublin$ is not even known to be close under the standard log-space reductions. Therefore, much weaker reductions may be more suitable to discuss the computational hardness of various real-life problems. A weaker notion, called ``short'' log-space reductions, was in fact invented and studied extensively in \cite{Yam18,Yam17a}. The importance of such reductions is exemplified by the fact that $\psublin$ is indeed closed under short log-space reductions.

\subsection{New Challenges in This Exposition}

The key question of whether LSH is true may hinge at the intensive study of parameterized-$\nl$ ``complete'' problems. It is however unfortunate that a very few parameterized decision problems have been proven to be equivalent in computational complexity to $(\mathrm{2SAT}_3,m_{vbl})$ by short log-space reductions, and this fact drives us to seek out new parameterized decision problems in this exposition in hope that we will eventually come to the point of proving the validity of LSH. This exposition is therefore devoted to proposing a new set of problems and proving their equivalence to $(\mathrm{2SAT}_3,m_{vbl})$ by appropriate short log-space reductions.

To replenish the existing short list of parameterized-$\nl$ complete problems, after reviewing fundamental notions and notation in Section \ref{sec:preparation}, we will propose three new decision problems in $\nl$, which are obtained by placing ``natural'' restrictions on instances of the following three
typical $\np$-complete combinatorial problems:
the \emph{vertex cover problem} (VC), the \emph{exact cover by 3-sets problem} (3XC), and the \emph{3-dimensional matching problem} (3DM) (refer to, e.g., \cite{GJ79} for their properties).
We will then set up their corresponding natural log-space size parameters to form the desired parameterized decision problems.

In Sections \ref{sec:vertex-cover}--\ref{sec:2-dimention}, we will prove that those new parameterized decision problems are equivalent in computational complexity to $(\mathrm{2SAT}_3,m_{vbl})$ by constructing appropriate short log-space reductions.
Since $(\mathrm{2SAT}_3,m_{vbl})$ is parameterized-$\nl$ complete, so are all the three new problems.
This completeness result immediately implies that, under the assumption of LSH, those problems cannot be solved in polynomial time using only sub-linear space.

\section{Fundamental Notions and Notation}\label{sec:preparation}

We briefly describe basic notions and notation necessary to read through the rest of this exposition.

\subsection{Numbers, Sets, Graphs, Languages, and Machines}\label{sec:number-graph}

We denote by $\nat$ the set of all \emph{natural numbers} including   $0$, and denote by $\integer$ the set of all \emph{integers}.
Let $\nat^{+}=\nat-\{0\}$. Given two numbers $m,n\in\integer$ with $m\leq n$, the notation $[m,n]_{\integer}$ expresses the  \emph{integer interval} $\{m,m+1,\ldots,n\}$. We further abbreviate $[1,n]_{\integer}$ as $[n]$ whenever $n\geq1$.
All \emph{polynomials} are assumed to take nonnegative coefficients and all \emph{logarithms} are taken to the base $2$.
The informal notion $polylog(n)$ refers to an arbitrary polynomial in $\log{n}$. Given a (column) vector $v=(a_1,a_2,\ldots,a_k)^T$ (where ``$T$'' is a transpose operator) and a number $i\in[k]$,
the notation $v_{(i)}$ indicates the $i$th entry $a_i$ of $v$.
For two (column) vectors $u$ and $v$ of dimension $n$, the notation $u\geq v$ means that the inequality $u_{(i)}\geq v_{(i)}$ holds for every index $i\in[n]$.
A \emph{$k$-set} refers to a set that consists of exactly $k$ distinct elements.

An \emph{alphabet} is a finite nonempty set of ``symbols'' or ``letters''. Given an alphabet $\Sigma$, a \emph{string} over $\Sigma$ is a finite sequence of symbols in $\Sigma$. The \emph{length} of a string $x$, denoted $|x|$, is the total number of symbols in $x$. The notation $\Sigma^*$ denotes the set of all strings over $\Sigma$. A \emph{language} over $\Sigma$ is a subset of $\Sigma^*$.

In this exposition, we will consider directed and undirected graphs and each graph is expressed as $(V,E)$ with a set $V$ of vertices and a set $E$ of edges. An edge between two vertices $u$ and $v$ in a directed graph is denoted by $(u,v)$, whereas the same edge in an undirected graph is denoted by $\{u,v\}$. Two vertices are called \emph{adjacent} if there is an edge between them.
When there is a path from vertex $u$ to
vertex $v$, we succinctly write $u\leadsto v$.
An edge of $G$ is said to be a \emph{grip} if its both endpoints have degree at most $2$.
Given a graph $G=(V,E)$, we set $m_{ver}(G)=|V|$ and $m_{edg}(G)=|E|$. The following property is immediate.

\begin{lemma}\label{graph-vertex-edge}
For any connected graph $G$ whose degree is at most $k$, it follows that $m_{ver}(G)\leq 2m_{edg}(G)$ and $m_{edg}(G)\leq k m_{ver}(G)/2$.
\end{lemma}

If a Boolean formula $\phi$ in the \emph{conjunctive normal form} (CNF) contains $n$ variables and $m$ clauses, then we set $m_{vbl}(\phi)=n$ and $m_{cls}(\phi)=m$ as two natural size parameters.
For later convenience, we call a literal $x$ in $\phi$ \emph{removable} if no clause in $\phi$ contains $\overline{x}$ (i.e., the negation of $x$).
We say that $\phi$ is \emph{in a clean shape} if each clause of $\phi$ consists of literals whose variables are all different. An \emph{exact} 2CNF Boolean formula has exactly two literals in each clause.

As a model of computation, we use the notions of \emph{multi-tape deterministic and nondeterministic Turing machines} (or DTMs and NTMs, for short), each of which is equipped with one read-only input tape, multiple rewritable work tapes, and (possibly) a write-once\footnote{A tape is \emph{write-once} if its tape head never moves to the left and it must move to the right whenever it writes a non-blank symbol.} output tape such that, initially, each input is written on the input tape surrounded by two endmarkers, $\cent$ (left endmarker) and $\dollar$ (right endmarker), and all the tape heads are stationed on the designated ``start cells.''
Given two alphabets $\Sigma_1$ and $\Sigma_2$, a function $f$ from $\Sigma_1^*$ to $\Sigma_2^*$ (resp., from $\Sigma_1^*$ to $\nat$) is \emph{computable in time $t(n)$ using $s(n)$ space} if there exists a DTM $M$ such that, on each input $x$, $M$ produces $f(x)$ (resp., $1^{f(x)}$) on the output tape before it halts within $t(|x|)$ steps with accesses to at most $s(|x|)$ work-tape cells (not counting the input-tape cells as well as the output-tape cells). We freely identify a decision problem with its corresponding language.
A decision problem is defined to be computable within time $t(n)$ using at most $s(n)$ space in a similar manner.

\subsection{Parameterized Decision Problems and Short Reductions}\label{sec:parameterized}

Throughout this exposition, we target decision problems (equivalently,  languages) that are \emph{parameterized} by size parameters,  which specify ``sizes'' (i.e., positive integers) of instances given to target problems and those sizes are used as a basis to measuring computational complexities (such as execution time and memory usage) of the problems.
More formally, for any input alphabet $\Sigma$, a \emph{size parameter} is a map from $\Sigma^*$ to $\nat^{+}$.
The information on the instance size is frequently used in solving problems, and thus it is natural to assume the easy ``computability'' of the size. A size parameter $m:\Sigma^*\to\nat$ is said to be \emph{log-space computable} if it is computable using $O(\log{|x|})$ space, where $x$ is a symbolic input.
A \emph{parameterized decision problem} is a pair $(A,m)$ with a language  $A$ over a certain alphabet $\Sigma$ and a size parameter $m$ mapping $\Sigma^*$ to $\nat^{+}$.

For any parameterized decision problem $(A,m)$, we say that \emph{$(A,m)$ is computable in polynomial time using sub-linear space} if there exists a DTM that solves $A$ in time polynomial in $|x|m(x)$ using  $O(m(x)^{1-\varepsilon}polylog(|x|))$  space, where $\varepsilon$ is a certain fixed constant in the real interval $(0,1]$.
A parameterized decision problem $(A,m)$ with log-space size parameter $m$ is in $\psublin$ if $(A,m)$ is computable in polynomial time using sub-linear space.


To discuss sub-linear-space computability, however, the standard log-space many-one reductions (or $\dl$-m-reductions, for short) are no longer useful. For instance,
it is unknown that all NL-complete problems parameterized by natural log-space size parameters are \emph{equally}  difficult to solve in polynomial time using sub-linear space. This is because $\psublin$ is not yet known to be closed under standard $\dl$-m-reductions.
Fortunately, $\psublin$ is proven to be closed under slightly weaker reductions, called ``short'' reductions \cite{Yam17a,Yam18}.

The short $\dl$-m-reducibility between two parameterized decision problems $(P_1,m_1)$ and $(P_2,m_2)$ is given as follows: $(P_1,m_1)$ is {\em short L-m-reducible to} $(P_2,m_2)$, denoted by $(P_1,m_1)\sLreduces(P_2,m_2)$, if there is  a polynomial-time, logarithmic-space computable function $f$ (which is called a \emph{reduction function}) and two constants $k_1,k_2>0$ such that, for any input string $x$,  (i) $x\in P_1$ iff $f(x)\in P_2$ and (ii) $m_2(f(x))\leq k_1 m_1(x)+k_2$.
Instead of using $f$, if we use a polynomial-time logarithmic-space oracle Turing machine $M$ to reduce $(P_1,m_1)$ to $(P_2,m_2)$ with the extra requirement of  $m_2(z)\leq k_1m_1(x)+k_2$ for any query word $z$ made by $M$ on input $x$ for oracle $P_2$, then $(P_1,m_1)$ is said to be \emph{short $\dl$-T-reducible to} $(P_2,m_2)$, denoted by $(P_1,m_1)\sLTreduces(P_2,m_2)$.

For any reduction $\leq$ in $\{\sLreduces, \sLTreduces\}$, we say that two parameterized decision problems  $(P_1,m_1)$ and $(P_2,m_2)$ are \emph{inter-reducible} (to each other) by $\leq$-reductions if both $(P_1,m_1)\leq(P_2,m_2)$ and $(P_2,m_2)\leq(P_1,m_1)$ hold; in this case, we briefly write $(P_1,m_1)\equiv(P_2,m_2)$.

\begin{lemma}{\rm \cite{Yam17a}}\label{transitivity}
Let $(L_1,m_1)$ and $(L_2,m_2)$ be two arbitrary parameterized decision problems. (1) If $(L_1,m_1)\sLreduces (L_2,m_2)$, then $(L_1,m_1)\sLTreduces (L_2,m_2)$. (2) If $(L_1,m_1)\sLTreduces (L_2,m_2)$ and $(L_2,m_2)\in \psublin$, then $(L_1,m_1)\in\psublin$.
\end{lemma}

\subsection{The Linear Space Hypothesis or LSH}

One of the first problems that were proven to be $\np$-complete in the past literature is the \emph{$3$CNF Boolean formula satisfiability problem} (3SAT), which asks whether or not a given 3CNF Boolean formula $\phi$ is satisfiable \cite{Coo71}.
In sharp comparison, its natural variant, called the \emph{$2$CNF Boolean formula satisfiability problem} ($\mathrm{2SAT}$), was proven to be $\nl$-complete \cite{JLL76} (together with the results of \cite{Imm88,Sze88}).
Let us further consider its natural restriction introduced in \cite{Yam17a}. Let $k\geq2$.

\s
{\sc $k$-Bounded $2$CNF Boolean Formula Satisfiability Problem}  ({\sc $2$SAT$_{k}$}):
\renewcommand{\labelitemi}{$\circ$}
\begin{itemize}\vs{-1}
  \setlength{\topsep}{-2mm}%
  \setlength{\itemsep}{1mm}%
  \setlength{\parskip}{0cm}%

\item {\sc Instance:} a 2CNF Boolean formula $\phi$ whose variables occur at most $k$ times each in the form of literals.

\item {\sc Question:} is $\phi$ satisfiable?
\end{itemize}

As natural log-space size parameters for the decision problem $\mathrm{2SAT}_k$, we use the aforementioned size parameters $m_{vbl}(\phi)$ and $m_{cls}(\phi)$.

Unfortunately, not all $\nl$-complete problems are proven to be  inter-reducible to one another by short log-space reductions.
An example of $\nl$-complete problems that are known to be inter-reducible to $(\mathrm{2SAT}_3,m_{vbl})$ is a variant of the directed $s$-$t$-connectivity problem whose instance graphs have only vertices of degree at most $k$ ($k\dstcon$) for any number $k\geq3$.

\s
{\sc Degree-$k$ Directed $s$-$t$ Connectivity Problem}  ({\sc $k$DSTCON}):
\renewcommand{\labelitemi}{$\circ$}
\begin{itemize}\vs{-1}
  \setlength{\topsep}{-2mm}%
  \setlength{\itemsep}{1mm}%
  \setlength{\parskip}{0cm}%

\item {\sc Instance:} a directed graph $G=(V,E)$ of degree at most $k$  and two vertices $s,t\in V$

\item {\sc Question:} is there any simple path in $G$ from $s$ to $t$?
\end{itemize}

Given a graph $G$ with $n$ vertices and $m$ edges, we set $m_{ver}(\pair{G,s,t})=n$ and $m_{edg}(\pair{G,s,t})=m$ as natural  log-space size parameters.

\begin{lemma}\label{DSTCON-equiv-2SAT}{\rm \cite{Yam17a}}
Let $k\geq3$ be any integer. (1) $(\mathrm{2SAT}_k,m_{vbl})$ is inter-reducible to $(\mathrm{2SAT}_k,m_{cls})$ and also to $(\mathrm{2SAT}_3,m_{vbl})$ by short $\dl$-m-reductions.
(2) $(k\dstcon,m_{ver})$ is inter-reducible to $(k\dstcon,m_{edg})$ and further to $(3\dstcon,m_{ver})$ by short $\dl$-m-reductions.
(3) $(3\dstcon,m_{ver})$ is inter-reducible to $(\mathrm{2SAT}_3,m_{vbl})$ by short $\dl$-T-reductions.
\end{lemma}

Notice that we do not know whether we can replace short $\dl$-T-reductions in Lemma \ref{DSTCON-equiv-2SAT}(3) by short $\dl$-m-reductions. This exemplifies a subtle difference between $\leq^{\mathrm{sL}}_{T}$ and $\leq^{\mathrm{sL}}_{m}$.

\begin{definition}\label{def:LSH}
The \emph{linear space hypothesis} (LSH) asserts, as noted in Section \ref{sec:linear-space-hypothesis}, the insolvability of the specific parameterized decision problem $(\mathrm{2SAT}_3,m_{vbl})$ within polynomial time using only sub-linear space.
\end{definition}

In other words, LSH asserts that $(\mathrm{2SAT}_3,m_{vbl})\notin \psublin$. Note that, since $\psublin$ is closed under short $\dl$-T-reductions by Lemma \ref{transitivity}(2), if a parameterized decision problem $(A,m)$ satisfies $(A,m)\equiv^{\mathrm{sL}}_{T} (\mathrm{2SAT}_3,m_{vbl})$, we can freely replace $(\mathrm{2SAT}_3,m_{vbl})$
in the definition of LSH by $(A,m)$. The use of short $\dl$-T-reduction can be relaxed to a much weaker notion of \emph{short $\mathrm{SLRF}$-T-reduction} \cite{Yam18,Yam17a}.

\subsection{Linear Programming and Linear Equations}\label{sec:LP-2LP-LE}

As a basis of later $\nl$-completeness proofs, we recall a combinatorial problem of Jones, Lien, and Laaser \cite{JLL76}, who studied a
problem of determining whether or not there exists a $\{0,1\}$-solution to a given set of linear programs, provided that each linear program (i.e., a linear inequality) has at most $2$ nonzero coefficients.
When each variable further has at most $k$ nonzero coefficients in all the linear programs, the corresponding problem is called the \emph{$(2,k)$-entry $\{0,1\}$-linear programming problem} \cite{Yam17a}, which is formally described as below.

\s
{\sc $(2,k)$-Entry $\{0,1\}$-Linear Programming Problem}  ({\sc LP$_{2,k}$}):
\renewcommand{\labelitemi}{$\circ$}
\begin{itemize}\vs{-1}
  \setlength{\topsep}{-2mm}%
  \setlength{\itemsep}{1mm}%
  \setlength{\parskip}{0cm}%

\item {\sc Instance:} a rational $m\times n$ matrix $A$, and a rational (column) vector $b$ of dimension $n$, where each row of $A$ has at most $2$ nonzero entries and each column has at most $k$ nonzero entries.

\item {\sc Question:} is there any $\{0,1\}$-vector $x$ for which $Ax\geq b$?
\end{itemize}

For practicality, all entries in $A$ are assumed to be expressed in binary using $O(\log{n})$ bits. For any instance $x$ of the form $\pair{A,b}$ given to $\mathrm{LP}_{2,k}$, we use two log-space size parameters defined as
$m_{row}(x)=n$ and $m_{col}(x)=m$.

It is known that, for any index $k\geq3$, the parameterized decision problem $(\mathrm{LP}_{2,k},m_{row})$ is  inter-reducible to $(\mathrm{LP}_{2,k},m_{col})$ and further to $(\mathrm{2SAT}_3,m_{vbl})$ by short $\dl$-m-reductions \cite{Yam17a}.

\begin{lemma}\label{PSLE-reduction}{\rm \cite{Yam17a}}
The following parameterized decision problems are all inter-reducible to one another by short $\dl$-m-reductions:
$(\mathrm{LP}_{2,k},m_{row})$, $(\mathrm{LP}_{2,k},m_{col})$, and
$(\mathrm{2SAT}_{3},m_{vbl})$
for every index $k\geq3$.
\end{lemma}


We can strengthen the requirement of the form $Ax\geq b$ in $\mathrm{LP}_{2,k}$ as follows. Consider another variant of $\mathrm{LP}_{2,k}$, in which we ask whether or not $b_1\geq Ax \geq b_2$ holds for a certain $\{0,1\}$-vector $x$ for any given matrix $A$ and two (column) vectors $b_1$ and $b_2$.

\s
{\sc Bidirectional $(2,k)$-Entry $\{0,1\}$-Linear Programming Problem}  ({\sc $2$LP$_{2,k}$}):
\renewcommand{\labelitemi}{$\circ$}
\begin{itemize}\vs{-1}
  \setlength{\topsep}{-2mm}%
  \setlength{\itemsep}{1mm}%
  \setlength{\parskip}{0cm}%

\item {\sc Instance:} a rational $m\times n$ matrix $A$, and two rational vectors $b_1$ and $b_2$ of dimension $n$, where each row of $A$ has at most $2$ nonzero entries and each column has at most $k$ nonzero entries.

\item {\sc Question:} is there any $\{0,1\}$-vector $x$ for which $b_1\geq Ax\geq b_2$?
\end{itemize}

\begin{proposition}\label{LP-vs-2LP}
For any index $k\geq3$, $(\mathrm{2LP}_{2,k},m_{col}) \equiv^{\mathrm{sL}}_{m} (\mathrm{LP}_{2,k},m_{col})$.
\end{proposition}

\begin{yproof}
The reduction $(\mathrm{LP}_{2,k},m_{col}) \leq^{\mathrm{sL}}_{m} (\mathrm{2LP}_{2,k},m_{col})$ is easy to verify by setting $b_2=b$ and $b_1=(b'_{i})_{i}$ with $b'_{i}=\max\{|a_{ij_1}|+|a_{ij_2}|: j_1,j_2\in[m],j_1<j_2\}$ for any instance pair $A=(a_{ij})_{ij}$ and $b=(b_{j})_{j}$ given to $\mathrm{LP}_{2,k}$.
Since the description size of $(A,b_1,b_2)$ is proportional to that of $(A,b)$, the reduction is indeed ``short.''

To verify the opposite reducibility $(\mathrm{2LP}_{2,k},m_{col}) \leq^{\mathrm{sL}}_{m} (\mathrm{LP}_{2,k+2},m_{col})$, it suffices to prove that $(\mathrm{2LP}_{2,k},m_{col}) \sLreduces  (\mathrm{LP}_{2,k},m_{col})$
since $(\mathrm{LP}_{2,l},m_{col}) \equiv^{\mathrm{sL}}_{m} (\mathrm{LP}_{2,3},m_{col})$ for any $l\geq3$ by Lemma \ref{PSLE-reduction}.
Take an arbitrary instance $(A,b,b')$ given to $\mathrm{2LP}_{2,k}$  and assume that $A=(a_{ij})_{ij}$ is an $m\times n$ matrix and $b=(b_i)_{i}$ and $b'=(b'_i)_{i}$ are two (column) vectors of dimension $m$.
We wish to reduce $(A,b,b')$ to an appropriate instance $(D,c)$ for  $\mathrm{LP}_{2,k}$, where $D=(d_{ij})_{ij}$ is a $4m\times 2n$ matrix and $c=(c_i)_{i}$ is a $4m$-dimensional vector. For all index pairs $i\in[m]$ and $j\in[n]$, let $d_{ij}=a_{ij}$, $d_{m+i,n+j}=-a_{ij}$, and $d_{i,n+j}=d_{m+i,j}=0$.
For all indices $i\in[m]$, let $c_i=b_i$  and $c_{m+i}=-b'_i$.
Moreover, for any pair $(i,j)\in[m]\times[n]$, we set $d_{2m+i,j} = 1$, $d_{2m+i,n+j} = -1$, and $c_{2m+i}=0$. In addition, we set $d_{3m+i,j}=-1$, $d_{3m+i,n+j}=1$, and $c_{3m+i}=0$. Notice that each column of $N$ has at most $k+2$ nonzero entries and each row of $D$ has at most $2$ nonzero entries.

Let $x=(x_j)_j$ denote a $\{0,1\}$-vector of dimension $n$ for $A$ and let  $y=(y_j)_j$ denote a $\{0,1\}$-vector of dimension $2n$ for $D$ satisfying $y_j=x_j$ and $y_{n+j}=x_j$ for any $j\in[n]$. It then follows that the inequality $\sum_{j=1}^{n}d_{ij}y_j \geq c_i$ is equivalent to $\sum_{j=1}^{n}a_{ij}x_j\geq b_j$. Furthermore, $\sum_{j=1}^{n}d_{n+i,j}y_{n+j}\geq c_{n+i}$ is equivalent to $-\sum_{j=1}^{n}a_{ij}x_j\geq -b'_j$, which is the same as $\sum_{j=1}^{n}a_{ij}x_j\leq b'_i$.
Therefore, we conclude that $b'\geq Ax\geq b$ iff $Dy\geq c$. In other words, it follows that $(A,b,b')\in \mathrm{2LP}_{2,k},$ iff $(D,c)\in \mathrm{LP}_{2,k}$.
\end{yproof}

As a special case of $2\mathrm{LP}_{2,k}$ by restricting its instances on the form $(A,b_1,b_2)$ with $b_1=b_2$, it is possible to consider the decision problem of asking whether or not $Ax =b$ holds for an appropriately chosen  $\{0,1\}$-vector $x$. We call this new problem the \emph{$(2,k)$-entry $\{0,1\}$-linear equation problem} ($\mathrm{LE}_{2,k}$).
As shown in Lemma \ref{LE-in-L}, $\mathrm{LE}_{2,k}$ falls into $\dl$, and thus this fact signifies how narrow the gap between $\nl$ and $\dl$ is.
For the proof of the lemma, we define the \emph{exclusive-or clause} (or the \emph{$\oplus$-clause}) of two literals $x$ and $y$ to be the formula  $x\oplus y$. The problem $\oplus\mathrm{2SAT}$ asks whether,  for a given collection $C$ of $\oplus$-clauses, there exists a truth assignment $\sigma$ that forces all $\oplus$-clauses in $C$ to be true. It is known that $\oplus\mathrm{2SAT}$ is in $\dl$ \cite{JLL76}.

\begin{lemma}\label{LE-in-L}
For any index $k\geq3$, $\mathrm{LE}_{2,k}$ belongs to $\dl$.
\end{lemma}

\begin{yproof}
Consider any instance $(A,b)$ given to $\mathrm{LE}_{2,k}$. Since $\oplus\mathrm{2SAT}\in \dl$, it suffices to reduce $\mathrm{LE}_{2,k}$ to $\oplus\mathrm{2SAT}$ by standard $\dl$-m-reductions.
Note that the equation $Ax=b$ is equivalent to $a_{ij_1}x_{j_1}+a_{ij_2}x_{j_2}=b_i$ for all $i\in[m]$, where $a_{ij_1}$ and $a_{ij_2}$ are nonzero entries of $A$ with $j_1,j_2\in[n]$.
Fix an index $i\in[m]$ and consider the first case where $j_1=j_2$. In this case, we translate $a_{ij_1}x_{j_1}=b_i$ into a $\oplus$-clause $v_{j_1}\oplus 0$ if $x_{ij_1}=1$, and $v_{j_1}\oplus 1$ otherwise. In the other case of $j_1\neq j_2$, on the contrary, we translate $a_{ij_1}x_{j_1}+a_{ij_2}x_{j_2}=b_i$ into two $\oplus$-clauses $\{x_{j_1}\oplus 0, x_{j_2}\oplus 1\}$ if $(x_{j_1},x_{j_2})=(1,0)$, and the other values of $(x_{j_1},x_{j_2})$ are similarly treated. Finally, we define $C$ to be the collection of all $\oplus$-clauses obtained by the aforementioned translations. It then follows that $Ax=b$ iff $C$ is satisfiable.
This implies that $(A,b)\in\mathrm{LE}_{2,k}$ iff $C\in\oplus\mathrm{2SAT}$.
\end{yproof}

\section{2-Checkered Vertex Covers}\label{sec:vertex-cover}

The \emph{vertex cover problem} (VC) is a typical $\np$-complete problem, which has been used as a basis of the completeness proofs of many other $\np$ problems, including the \emph{clique problem} and the \emph{independent set problem} (see, e.g., \cite{Kar72,GJ79}).
For a given undirected graph $G=(V,E)$, a \emph{vertex cover} for $G$  is a subset $V'$ of $V$ such that, for each edge $\{u,v\}\in E$, at least one of the endpoints $u$ and $v$ belongs to $V'$.

The problem VC remains $\np$-complete even if its instances are limited to planar graphs. Similarly, the vertex cover problem restricted to graphs of degree at least 3 is also $\np$-complete; however, the same problem falls into $\dl$ if we require graphs
to have degree at most $2$.
We wish to seek out a reasonable setting situated between those two special cases.
For this purpose, we intend to partition all edges into two categories: grips and non-grips (where ``grips'' are defined in Section \ref{sec:number-graph}).
Since grips have a simpler structure than non-grips, the grips need to be treated slight differently from the others.
In particular, we request an additional condition, called 2-checkeredness, which is described as follows.
A subset $V'$ of $V$ is called \emph{2-checkered} exactly when, for any edge $e\in E$, if both endpoints of $e$ are in $V'$, then $e$ must be a grip.
The \emph{2-checkered vertex cover problem} is introduced in the following way.

\s
{\sc 2-Checkered Vertex Cover Problem} ({\sc 2CVC}):
\renewcommand{\labelitemi}{$\circ$}
\begin{itemize}\vs{-1}
  \setlength{\topsep}{-2mm}%
  \setlength{\itemsep}{1mm}%
  \setlength{\parskip}{0cm}%

\item {\sc Instance:} an undirected graph  $G=(V,E)$.

\item {\sc Question:} is there a 2-checkered vertex cover for $G$?
\end{itemize}

Associated with the decision problem $\mathrm{2CVC}$, we set up the log-space  size parameters:  $m_{ver}(G)$ and $m_{edg}(G)$, which respectively express the total number of the vertices of $G$ and that of the edges of $G$.

Given an instance of graph $G=(V,E)$ to $\mathrm{2CVC}$, if we further demand that every vertex in $V$ should have degree at most $k$ for any fixed constant $k\geq3$, then we denote by $\mathrm{2CVC}_{k}$ (Degree-$k$ 2CVC) the problem obtained from $\mathrm{2CVC}$.
There exists a close connection between the parameterizations of $\mathrm{2CVC}_{3}$ and  $\mathrm{2SAT}_{3}$.

\begin{theorem}\label{2CVC-LP}
$(\mathrm{2CVC}_{3},m_{ver}) \equiv^{\mathrm{sL}}_{m} (\mathrm{2CVC}_{3},m_{edg}) \equiv^{\mathrm{sL}}_{m} (\mathrm{2SAT}_{3},m_{vbl})$.
\end{theorem}


\begin{yproof}
Firstly, it is not difficult to show that  $(\mathrm{2CVC}_{3},m_{ver}) \equiv^{\mathrm{sL}}_{m} (\mathrm{2CVC}_{3},m_{edg})$ by Lemma \ref{graph-vertex-edge}.

Next, we intend to prove that $(2\mathrm{SAT}_3,m_{vbl})\leq^{\mathrm{sL}}_{m} (\mathrm{2CVC}_{3},m_{ver})$. Let $\phi$ be any instance to $\mathrm{2SAT}_3$ made up of a set $U=\{u_1,u_2,\ldots,u_n\}$ of  variables and a set $C=\{c_1,c_2,\ldots,c_m\}$ of 2CNF Boolean clauses.
For convenience, we write  $\overline{U}$ for the set $\{\overline{u_1},\overline{u_2},\ldots,\overline{u_n}\}$ of negated variables and define $\hat{U}=U\cup\overline{U}$.
In the case where a clause contains any removable literal $x$, it is possible to delete all clauses that contain $x$, because we can freely assign $T$ (true) to $x$. Without loss of generality, we  assume that there is no removable literal in $\phi$.
We further assume that $\phi$ is an exact 2CNF formula in a clean shape (explained in Section \ref{sec:number-graph}). Since every clause has exactly two literals, each clause $c_j$ is assumed to have the form $c_{j}[1]\vee c_{j}[2]$ for any index $j\in[m]$, where $c_{j}[1]$ and $c_{j}[2]$ are treated as ``labels'' that \emph{represent} two literals in the clause $c_j$.

Let us construct an undirected graph $G=(V,E)$ as follows.
We define $V=\{u_i^{(1)},u_i^{(2)}, c_{j}[1],c_{j}[2] \mid i\in[n],j\in[m]\}$ and we set $\tilde{U}$ to be $\{u_i^{(1)},u_i^{(2)}\mid i\in[n]\}$ by writing $u_i^{(1)}$ for  $u_i$ and $u_i^{(2)}$ for  $\overline{u_i}$.  We further set
$E$ as the union of  $\{ \{u_i^{(1)},u_i^{(2)}\}, \{c_{j}[1],c_{j}[2]\} \mid i\in[n], j\in[m]\}$ and $\{ \{z,c_{j}[l]\}\mid \text{ $z\in \tilde{U}$, $l\in[2]$, and $c_{j}[l]$ represents $z$ }\}$.
Since each clause contains exactly two literals, it follows that $deg(c_{j}[1])=deg(c_{j}[2])=2$.
Thus, the edge $\{c_j[1],c_j[2]\}$ for each index $j$ is  a grip.  Moreover, since each variable $u_i$ appears at most $3$ times in the form of literals (because of the condition of $\mathrm{2SAT}_3$), $deg(u_i^{(1)})+deg(u_i^{(2)})\leq 5$.
Since no removable literal exists in $\phi$, we obtain $\max\{deg(u_i^{(1)}), deg(u_i^{(2)})\}\leq 3$.
It follows by the definition that $m_{ver}(G)=2(|U|+|C|)\leq 8|U|=8m_{vbl}(\phi)$ since $|C|\leq 3|U|$.

Here, we want to verify that $\phi\in\mathrm{2SAT}_3$ iff $G\in\mathrm{2CVC}_3$. Assume that $\phi\in\mathrm{2SAT}_3$. Let $\sigma: U\to \{T,F\}$ be any truth assignment that makes $\phi$ satisfiable.
We naturally extend $\sigma$ to a map from $\hat{U}$ to $\{T,F\}$ by setting $\sigma(\bar{u})$ to be the opposite of $\sigma(u)$.  Its corresponding  vertex cover $C_{\sigma}$ is defined in two steps. Initially, $C_{\sigma}$ contains all elements $z\in \hat{U}$ satisfying $\sigma(z)= F$.
For each index  $j\in[m]$, let $A_j =\{i\in[2]\mid \exists z\in \hat{U} [\text{ $c_j[i]$ represents $z$ and $\sigma(z)=T$ }] \}$.
Notice that $A_j\subseteq \{1,2\}$.
If $A_j=\{i\}$ for a certain $i\in[2]$, then we append to $C_{\sigma}$ the vertex $c_j[i]$; however, if $A_j=\{1,2\}$, then we append to $C_{\sigma}$ the two vertices $c_j[1]$ and $c_j[2]$ instead.

To illustrate our construction, let us consider a simple example: $\phi\equiv c_1\wedge c_2\wedge c_3\wedge c_4$ with $c_1\equiv u_1\vee \overline{u_2}$, $c_2\equiv u_2\vee u_1$,  $c_3\equiv \overline{u_1}\vee u_3$, and $c_4\equiv u_2\vee \overline{u_3}$. The corresponding graph $G$ is drawn in Fig.~\ref{fig:2CVC-graph-figure}. Take the truth assignment $\sigma$ that satisfies $\sigma(u_1)=\sigma(u_2)=\sigma(x_3)=T$. We then obtain $A_1=\{1\}$, $A_2=\{1,2\}$, and $A_3=\{2\}$. Therefore, the resulting 2-checkered vertex cover $C_{\sigma}$ is the set $\{u_1^{(2)}, u_2^{(2)}, u_3^{(2)}, c_1[1], c_2[1], c_2[2], c_3[2], c_4[1] \}$.

\begin{figure}[t]
\centering
\includegraphics*[height=2.5cm]{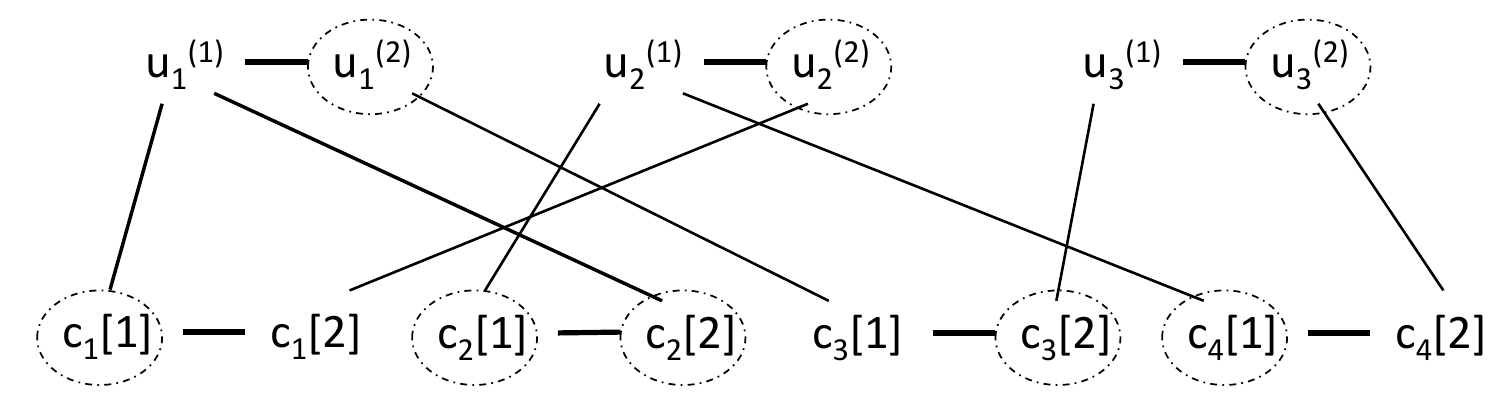}
\caption{The graph $G$ obtained from $\phi\equiv c_1\wedge c_2\wedge c_3\wedge c_4$ with  $c_1\equiv u_1\vee \overline{u_2}$, $c_2\equiv u_2\vee u_1$,  $c_3\equiv \overline{u_1}\vee u_3$, and $c_4\equiv u_2\vee \overline{u_3}$. For the truth assignment  $\sigma$ satisfying  $\sigma(x_1)=\sigma(x_2)=\sigma(x_3)=T$, the 2-checkered vertex cover $C_{\sigma}$ consists of all vertices marked by dotted circles.}\label{fig:2CVC-graph-figure}
\end{figure}

By the definition of $C_{\sigma}$'s, we conclude that $G$ belongs to $\mathrm{2CVC}_3$. Conversely, we assume that $\phi\notin\mathrm{2SAT}_3$. Consider any truth assignment $\sigma$ for $\phi$ and construct $C_{\sigma}$ as before. By the construction of $C_{\sigma}$, if $C_{\sigma}$ is a 2-checkered vertex cover, then $\sigma$ should force $\phi$ to be true, a contradiction. Hence, $G\notin \mathrm{2CVC}_3$ follows. Overall,
it follows that
$\phi\in \mathrm{2SAT}_3$ iff $G\in \mathrm{2CVC}_3$. Therefore, we obtain $(2\mathrm{SAT}_3,m_{vbl})\leq^{\mathrm{sL}}_{m} (\mathrm{2CVC}_{3},m_{ver})$.

Conversely, we need to prove that $(\mathrm{2CVC}_{3},m_{ver}) \leq^{\mathrm{sL}}_{m} (2\mathrm{SAT}_3,m_{vbl})$.
Given an undirected graph $G=(V,E)$, we want to define a 2CNF Boolean formula $\phi$ to which $G$ reduces.
Let $V=\{v_1,v_2,\ldots,v_n\}$ and $E=\{e_1,e_2,\ldots,e_m\}$ for certain numbers $m,n\in\nat^{+}$.

Hereafter, we use the following abbreviation: $u\rightarrow v$ for $\overline{u}\vee v$, $u\leftrightarrow v$ for $(u\rightarrow v)\wedge (v \rightarrow u)$, and $u\not\leftrightarrow v$ for $(u\vee v)\wedge (\overline{u}\vee \overline{v})$. Notice that,
as the notation $\not\leftrightarrow$ itself suggests,
$u \not\leftrightarrow v$ is logically equivalent to the negation of $u\leftrightarrow v$.

We first define a set $U$ of variables to be $V$. For each edge $e=\{u,v\}\in E$, we define $C_e$ as follows. If one of $u$ and $v$ has degree more than $2$,  then we set $C_e$ to be $u\not\leftrightarrow v$; otherwise, we set $C_e$ to be $u\vee v$. Finally, we define $C$ to be the set of all clauses, namely, $\{C_e\mid e\in E\}$. Let $\phi$ denote the 2CNF Boolean formula made up of all clauses in $C$.

Next, we intend to verify that $G$ has a 2-checkered vertex cover iff $\phi$ is satisfiable. Assume that $G$ has a 2-checkered vertex cover, say, $V'$. Consider $C$ obtained from $G$. We define a truth assignment $\sigma$ by setting $\sigma(v)=T$ iff $v\in V'$.
Take any edge $e=\{u,v\}$. If one of $u$ and $v$ has degree more than $2$, then either ($u\in V'$ and $v\notin V'$) or ($u\notin V'$ and $v\in V')$ hold, and thus $\sigma$ forces $u\not\leftrightarrow v$ to be true.
Otherwise, since either $u\in V'$ or $v\in V'$, $\sigma$ forces $u\vee v$ to be true. This concludes that $\phi$ is satisfiable.
On the contrary, we assume that $\phi$ is satisfiable by a certain truth assignment, say, $\sigma$; that is, for any edge $e\in E$, $\sigma$ forces $C_e$ to be true. We define a subset $V'$ of $V$ as $V'=\{v\in V\mid \sigma(v)=T\}$. Let $e=\{u,v\}$ be any edge. If $C_e$ has the form $u\vee v$ for $u,v\in V$, then either $u$ or $v$ should belong to $V'$. If $\sigma$ forces $u\not\leftrightarrow v$ in $C$ to be true, then either ($u\in V'$ and $v\notin V'$) or ($u\notin V'$ and $v\in V'$) hold. Hence, $V'$ is a 2-checkered vertex cover.
\end{yproof}


The $\nl$-completeness of $\mathrm{2CVC}_3$ follows from Theorem \ref{2CVC-LP} since $\mathrm{2SAT}$ (and also $\mathrm{2SAT}_3$) is $\nl$-complete by standard $\dl$-m-reductions \cite{JLL76} (based on the fact $\nl=\co\nl$ \cite{Imm88,Sze88}).

As an immediate corollary of Theorem \ref{2CVC-LP}, we obtain the following hardness result regarding the computational complexity of  $(\mathrm{2CVC},m_{ver})$ under the assumption of LSH.

\begin{corollary}\label{LSH-broomHVC}
Under LSH, letting $\varepsilon$ be any constant in $(0,1]$, there is no polynomial-time algorithm that solves $(\mathrm{2CVC},m_{ver})$ using $O(m_{ver}(x)^{1-\varepsilon})$ space, where $x$ is a symbolic input.
\end{corollary}


\begin{yproof}
Assume that LSH is true. If $(\mathrm{2CVC},m_{ver})$ is solvable in polynomial time using $O(m_{ver}(x)^{1-\varepsilon})$ space for a certain constant $\varepsilon\in(0,1)$, since $\mathrm{2CVC}_3$ is a ``natural'' subproblem of $\mathrm{2CVC}$, Theorem \ref{2CVC-LP} implies the existence of a polynomial-time algorithm that solves $(\mathrm{2SAT}_{3},m_{vbl})$ using $O(m_{row}(x)^{1-\varepsilon})$ space as well. This implies that LSH is false, a contradiction.
\end{yproof}

\section{Exact Covers with Exemption}\label{sec:exact-cover}

The \emph{exact cover by 3-sets problem} (3XC) was shown to be $\np$-complete \cite{Kar72}.
Fixing a \emph{universe} $X$, let us choose a collection $C$ of subsets of $X$. We say that $C$ is a \emph{set cover for $X$}  if every element in $X$ is contained in a certain set in $C$. Furthermore, given a subset $R\subseteq X$, $C$ is said to be an \emph{exact cover for $X$ exempt from} $R$ if (i) every element in $X-R$ is contained in a unique member of $C$ and (ii) every element in $R$ appears in at most one member of $C$. When $R=\setempty$, we say that $C$ is an \emph{exact cover for} $X$. Notice that any exact cover with exemption is a special case of a set cover.

To obtain a decision problem in $\nl$, we need one more restriction. Given a collection $C\subseteq \PP(X)$, we introduce a measure, called ``overlapping cost,'' of an element of any set in $C$ as follows.
For any element $u\in X$, the \emph{overlapping cost of $u$ with respect to (w..r.t.) $C$} is the cardinality $|\{A\in C \mid u\in A\}|$.
With the use of this special measure, we define the notion of $k$-overlappingness for any $k\geq2$ as follows. We say that $C$ is \emph{$k$-overlapping} if the overlapping cost of every element $u$ in $X$ w.r.t. $C$ is at most $k$.

\s
{\sc 2-Overlapping Exact Cover by $k$-Sets with Exemption Problem} ({\sc $k$XCE}$_2$):
\renewcommand{\labelitemi}{$\circ$}
\begin{itemize}\vs{-1}
  \setlength{\topsep}{-2mm}%
  \setlength{\itemsep}{1mm}%
  \setlength{\parskip}{0cm}%

\item {\sc Instance:} a finite set $X$, a subset $R$ of $X$,  and a $2$-overlapping collection $C$ of subsets of $X$ such that each set in $C$ has at most $k$ elements.

\item {\sc Question:} does $C$ contain an exact cover for $X$ exempt from $R$?
\end{itemize}

The use of an exemption set $R$ in the above definition is crucial. If we are given a 2-overlapping family $C$ of subsets of $X$ as an input and then ask for the existence of an exact cover for $X$, then the corresponding problem is rather easy to solve in log space \cite{AG00}.

The size parameter $m_{set}$ for $k\mathrm{XCE}_2$ satisfies $m_{set}(\pair{X,R,C}) = |C|$, provided that all elements of $X$ are expressed in $O(\log|X|)$ binary symbols. Obviously, $m_{set}$ is a log-space size parameter. In what follows, we consider $\mathrm{3XCE}_2$ parameterized by $m_{set}$, $(\mathrm{3XCE}_2,m_{set})$, and prove its inter-reducibility to $(\mathrm{2SAT}_3,m_{vbl})$..

\begin{theorem}\label{3XCE-equiv-2SAT}
$(\mathrm{3XCE}_{2},m_{set}) \equiv^{\mathrm{sL}}_{m} (\mathrm{2SAT}_{3},m_{vbl})$.
\end{theorem}

Theorem \ref{3XCE-equiv-2SAT} immediately implies the $\nl$-completeness of $\mathrm{3XCE}$.
To simplify the following proof, we recall from Section \ref{sec:LP-2LP-LE} the $\nl$-problem $\mathrm{2LP}_{2,k}$ and the fact that, for any index $k\geq3$, $(\mathrm{2LP}_{2,k},m_{col})\equiv^{\mathrm{sL}}_{m} (\mathrm{2SAT}_{3},m_{vbl})$, obtained from Lemma \ref{PSLE-reduction} and Proposition \ref{LP-vs-2LP}.

\bs
\begin{yproof}
We begin our proof with verifying that $(\mathrm{2SAT}_{3},m_{vbl}) \leq^{\mathrm{sL}}_{m} (\mathrm{3XCE}_{2},m_{set})$.
Let $\phi$ denote a $2$CNF Boolean formula with $n$ variables and $m$ clauses, given as an instance to $\mathrm{2SAT}_3$. Let $V$ denote the set $\{x_1,x_2,\ldots,x_n\}$ of all distinct variables in $\phi$ and let $C$ denote the set $\{C_1,C_2,\ldots,C_m\}$ of all (distinct) clauses in $\phi$.
With no loss of generality, we assume that there is no removable literal in $\phi$ and that $\phi$ is an exact formula in a clean shape. We write $\overline{V}$ for the set $\{\overline{x_1},\overline{x_2},\ldots,\overline{x_n}\}$ and define $\hat{V} = V\cup\overline{V}$. We freely identify a clause of the form $z_{i_1}\vee z_{i_2}$ for literals $z_{i_1}$ and $z_{i_2}$ with the set $\{z_{i_1},z_{i_2}\}$, which is clearly a subset of $\hat{V}$. By our assumption, each variable $x_i$ should  appear at most 3 times in different clauses in the form of literals.

We want to reduce $\phi$ to an appropriately constructed instance $(X,R,D)$ of  $\mathrm{3XCE}_2$. To construct such an instance, we first define the following three sets $X_1$, $X_2$, and $X_3$:
$X_1 =\{x_i[j] \mid i\in[n], j\in[m], x_i\in C_j\} \cup \{\overline{x_i}[j] \mid i\in[n], j\in[m], \overline{x_i} \in C_j\}$, $X_2=\{ s_j\mid j\in[m]\}$, and $X_3= \{t_i[j] \mid i\in[n],j\in[3]\}$. The universe $X$ is made up from those three sets  (i.e., $X= X_1\cup X_2\cup X_3$).

To understand the following construction better, we here illustrate a simple example of  $\phi$, which is of the form $C_1\wedge C_2\wedge C_3\wedge C_4$ with clauses $C_1 =\{x_1, \overline{x_2}\}$, $C_2=\{ x_1, x_3\}$, $C_3=\{ x_2, \overline{x_3}\}$, and $C_4=\{ \overline{x_1},  \overline{x_3}\}$. We define the set $D$ as drawn in  Fig.~\ref{fig:3XCE-graph-figure}. Take a truth assignment $\sigma$ defined by  $\sigma(x_1)=\sigma(x_2)=T$ and $\sigma(x_3)=F$. The exact cover $X'_{\sigma}$ for $X$ exempt from $R$ consists of $\{x_1[1],s_1\}$, $\{x_1[2],s_2\}$, $\{x_2[3],s_3\}$, $\{\overline{x_3}[4],s_4\}$, $\{x_1[2],t_1[1],t_1[2]\}$, and  $\{\overline{x_3}[2],t_3[1],t_3[2]\}$.

\begin{figure}[t]
\centering
\includegraphics*[height=2.8cm]{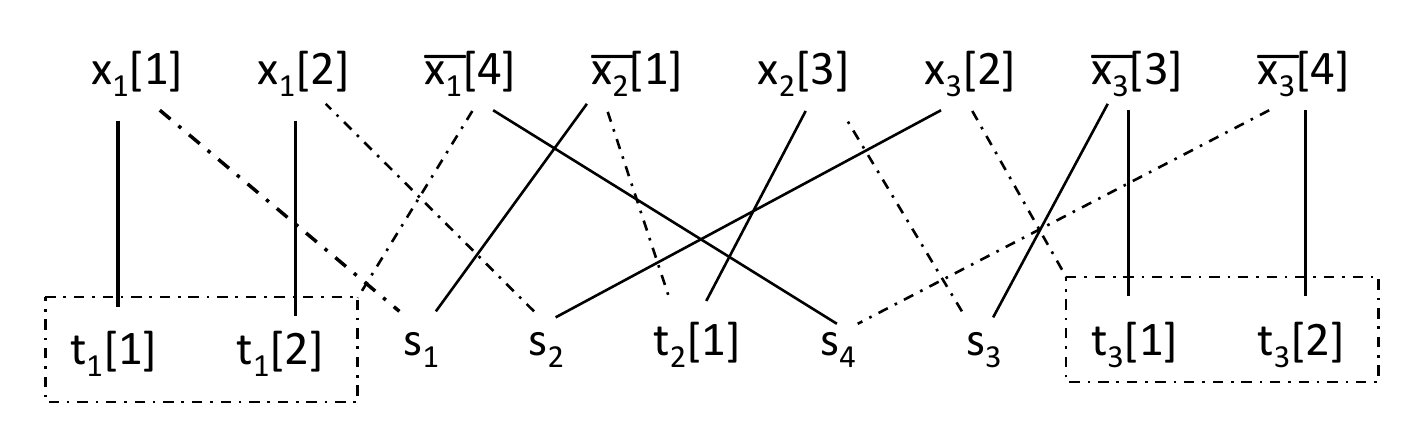}
\caption{The graph $G$ obtained from $\phi\equiv C_1\wedge C_2\wedge C_3\wedge C_4$ with clauses $C_1=\{ x_1, \overline{x_2}\}$, $C_2=\{ x_1, x_3\}$, $C_3= \{ x_2, \overline{x_3}\}$, and $C_4= \{ \overline{x_1}, \overline{x_3}\}$. The truth assignment $\sigma$ satisfies $\sigma(x_1)=\sigma(x_2)=T$ and $\sigma(x_3)=F$. The exact cover $X'_{\sigma}$ exempt from $R$, obtained from $\sigma$, consists of vertex pairs and triplets linked respectively by dotted lines and dotted boxes. Here, $t_1[3]$, $t_2[2]$, $t_2[3]$, and $t_3[3]$ are omitted for simplicity.}\label{fig:3XCE-graph-figure}
\end{figure}

Returning to the proof, let us define two groups of sets. For each index $j\in[m]$, $A_j$ is composed of the following 2-sets: $A_j=\{\{z_{i_1}[j],s_j\}, \{z_{i_2}[j],s_j\} \mid i_1,i_2\in[n], C_j=\{z_{i_1},z_{i_2}\}\subseteq \hat{V} \}$.
Associated with $V$, we set $V^{(+)}$ to be composed of all variables  $x_i$  such that $x_i$ appears in two clauses and $\overline{x_i}$ appears in one clause. Similarly, let $V^{(-)}$ be composed of all variables $x_i$ such that $x_i$ appears in one clause and $\overline{x_i}$ appears in two clauses. In addition, let $V^{(*)}$ consist of all other variables.
Note that, since there is no removable literal in $\phi$, any variable $x_i$ in $V^{(*)}$ appears in one clause and its negation $\overline{x_i}$ appears also in one clause.
Our assumption guarantees that $V=V^{(+)}\cup V^{(-)}\cup V^{(*)}$.
For each variable $x_i\in V^{(+)}$, we set $B_i^{(+)} =\{ \{x_i[j_1],t_i[1]\}, \{x_i[j_2],t_i[2]\}, \{\overline{x_i}[j_3],t_i[1],t_i[2]\} \}$, provided that $C_{j_1}$ and $C_{j_2}$ both contain $x_i$ and $C_{j_3}$ contains $\overline{x_i}$ for certain indices $j_1$, $j_2$, and $j_3$ with $j_1<j_2$.
Similarly, for each variable $x_i\in V^{(-)}$, we set $B_i^{(-)} =\{ \{\overline{x_i}[j_1],t_i[1]\}, \{\overline{x_i}[j_2],t_i[2]\}, \{x_i[j_3],t_i[1],t_i[2]\} \}$, provided that $C_{j_1}$ and $C_{j_2}$ both contain $\overline{x_i}$ and $C_{j_3}$ contains $x_i$.
In contrast, given any variable $x_i\in V^{(*)}$, we define $B_i^{(*)} = \{\{x_i[j_1],t_i[1]\}, \{\overline{x_i}[j_2],t_i[1]\}  \}$.
Finally, we set $D = (\bigcup_{j\in[m]}A_j) \cup (\bigcup_{i\in[n]} (B_i^{(+)}\cup B_i^{(-)}\cup B_i^{(*)}))$.
Notice that every element in $X$ is covered by exactly two sets in $D$.
To complete our construction, an exemption set $R$ is defined to be $X_1$.

Hereafter, we intend to verify that $\phi$ is satisfiable iff there exists an exact cover for $X$ exempt from $R$.
Given a truth assignment $\sigma:V\to\{T,F\}$, we define a set $X'_{\sigma}$ as follows. We first define $X'_1$ to be the set $\{ z[j],s_j \mid j\in[m], \sigma(z)=T, z\in \hat{V}\}$. For each element $x_i\in V^{(+)}$, if  $\sigma(x_i)=F$, then we set $X_{2,i}'=\{ \{x_i[j_1],t_i[1]\}, \{x_i[j_2],t_i[2]\} \} \subseteq B^{(+)}_i$, and if $\sigma(\overline{x_i})=F$, then we set $X_{2,i}'=\{ \{\overline{x_i}[j_3],t_i[1],t_i[2]\} \} \subseteq B^{(+)}_i$.
Similarly, for each element $x_i\in V^{(-)}$, we can define $X'_{2,i}$.
For any element $x_i\in V^{(*)}$, however,  if $\sigma(z)=F$ for a literal  $z\in\{x_i,\overline{x_i}\}$, then we define $X'_{2,i}= \{\{z[j_1],t_i[1]\}\}\subseteq B_i^{(*)}$.
In the end, $X'_{\sigma}$ is set to be the union $X'_1\cup(\bigcup_{i\in[n]}X'_{2,i})$.
Assume that $\phi$ is true by $\sigma$.
Since all clauses $C_j$ are true by $\sigma$, each $s_j$ in $X_2$ has overlapping cost of $1$ in $X'_{\sigma}$.
Moreover, each $t_i[j]$ in $X_3$ has overlapping cost of $1$ in $X'_{\sigma}$.
Either $x_i[j]$ or $\overline{x_i}[j]$ in $X_1$ has overlapping cost of at most $1$. Thus, $X'_{\sigma}$ is an exact cover for $X$ exempt from $R$ ($=X_1$).

On the contrary, we assume that $X'$ is an exact cover for $X$ exempt from $R$. We define a truth assignment $\sigma$ as follows.
For each 2-set $A_j$, if $\{z_{i_d}[j],s_j\}\in X'$ for a certain index $d\in[2]$, then we set $\sigma(z_{i_d})=T$. For each $B^{(+)}_i$, if $\{x_i[j_1],t_i[1]\},\{x_i[j_2],t_i[2]\}\in X'$, then we set $\sigma(x_i)=F$. If $\{\overline{x_i}[j_3],t_i[1],t_i[2]\}\in X'$, then we set $\sigma(\overline{x_i})=F$. The case of $B^{(-)}_i$ is similarly handled. In the case of $B_i^{(*)}$, if $\{z[j_1],t_i[1]\}\in X'$ for a certain $z\in\{x_i,\overline{x_i}\}$, then we set $\sigma(z)=F$.
Since $X'$ is an exact cover for $X-R$, for any clause $C_j$, there exists exactly one $z$ in $C_j$ satisfying $\sigma(z)=T$.


Conversely, we intend to verify that  $(\mathrm{3XCE}_{2},m_{set}) \leq^{\mathrm{sL}}_{m} (\mathrm{2SAT}_{3},m_{vbl})$.
Since $(\mathrm{2SAT}_{3},m_{vbl})\equiv^{\mathrm{sL}}_{m} (\mathrm{2LP}_{2,3},m_{row})$ by Lemma \ref{PSLE-reduction} and Proposition \ref{LP-vs-2LP}, if we show that   $(\mathrm{3XCE}_{2},m_{set}) \leq^{\mathrm{sL}}_{m} (\mathrm{2LP}_{2,3},m_{row})$, then we immediately obtain the desired consequence of $(\mathrm{3XCE}_2,m_{set})\sLreduces (\mathrm{2SAT}_3,m_{vbl})$.
Toward the claim of $(\mathrm{3XCE}_{2},m_{set}) \leq^{\mathrm{sL}}_{m} (\mathrm{2LP}_{2,3},m_{row})$, let us take an arbitrary instance $(X,R,C)$ given to $\mathrm{3XCE}_{2}$ with  $X=\{u_1,u_2,\ldots,u_n\}$ and $C=\{C_1,C_2,\ldots,C_m\}$. Notice that $R\subseteq X$ and $|C_i|\leq 3$ for all $i\in[m]$.

As the desired instance to $\mathrm{2LP}_{2,3}$, we define an $n\times m$ matrix $A=(a_{ij})_{ij}$ and two (column) vectors $b=(b_i)_{i}$ and $b'=(b'_i)_{i}$ as follows. Since each $u_i$ has overlapping cost of $2$, if $u_i$ is in $C_{j_1}\cap C_{j_2}$ for two distinct indices $j_1$ and $j_2$, then we set $a_{ij_1}=a_{ij_2}=1$. Let $b_i=b'_i=1$ for all $i\in[n]$ satisfying $u_i\in X-R$, and let $b_i=0$ and $b'_i=1$ for all $i\in[n]$ satisfying $u_i\in R$.
If $D$ is a set cover, then we define $x_{D}=(x_j)_j$ as follows: if $C_j\notin D$, then we set $x_j=1$; otherwise, we set $x_j=0$. We then want to show that $D$ is an exact cover for $X$ exempt from $R$ iff $x_{D}$ satisfies $b'\geq Ax_{D} \geq b$.
Assume that $D$ is an exact cover for $X$ exempt from $R$. Note that, if $u_i\in C_{j_1}\cap C_{j_2}$ for two distinct indices $j_1$ and $j_2$, then $\sum_{j=1}^{n}a_{ij}x_j = a_{ij_1}x_{j_1}+a_{ij_2}x_{j_2}=x_{j_1}+x_{j_2}$. Since $D$ contains exactly one set containing $u_i$, we obtain $x_{j_1}+x_{j_2}=1$. If $u_i\in R$, then we obtain $\sum_{j=1}^{n}a_{ij}x_j \in\{0,1\}$. Thus, we conclude that $b'\geq Ax_{D}\geq b$.
On the contrary, assume that $b'\geq Ax_{D}\geq b$. We obtain $\sum_{j=1}^{n}a_{ij}x_j = a_{ij_1}x_{j_1}+a_{ij_2}x_{j_2} = x_{j_1}+x_{j_2}$ for two indices $j_1$ and $j_2$ satisfying $a_{j_1}\neq0$ and $a_{j_2}\neq0$. If $u_i\notin R$, then $x_{j_1}+x_{j_2}=1$ holds because of $b_i=b'_i=1$, and thus exactly one of $C_{j_1}$ and $C_{j_2}$ must be in $D$. If $u_i\in R$, then $1\geq x_{j_1}+x_{j_2}\geq0$ holds, and thus at most one of $C_{j_1}$ and $C_{j_2}$ belongs to $D$. Therefore, $D$ is an exact cover for $X$ exempt from $R$.
\end{yproof}

Similarly to Corollary \ref{LSH-broomHVC}, we obtain the following statement concerning $\mathrm{3XCE}$.

\begin{corollary}\label{LSH-Steiner-tree}
Under LSH, no polynomial-time algorithm solves $(\mathrm{3XCE},m_{ver})$ using $O(m_{ver}(x)^{1-\varepsilon})$ space for a certain constant $\varepsilon\in(0,1)$, where $x$ is a symbolic input.
\end{corollary}

\section{Almost All Pairs 2-Dimensional Matching}\label{sec:2-dimention}

The \emph{3-dimensional matching problem} (3DM) is well-known to be $\np$-complete \cite{Kar72} while the 2-dimensional matching problem (2DM), which is seen as a bipartite perfect matching problem, falls into $\p$. In fact,  $\mathrm{2DM}$ has been proven to be $\nl$-hard \cite{CSV84} but it is not yet known to be in $\nl$. In this exposition, we wish to place our interest on a natural variant of $\mathrm{2DM}$, which turns to be $\nl$-complete.
Let us take a finite set $X$ and consider the Cartesian product $X\times X$.
For any two elements $(u,v),(w,z)\in X\times X$, we say that $(u,v)$ \emph{agrees with} $(w,z)$, denoted  $(u,v)\sqcap (w,z)\neq\emptyset$, if either $u=w$ or $v=z$.
A \emph{matching} over $X\times X$ is a subset $M$ of $X\times X$ such that no two distinct elements in $M$ agree with each other.
Given a subset $M \subseteq X\times X$, we define $M_{(1)} = \{u\in X \mid \exists v[(u,v)\in M]\}$
and $M_{(2)} = \{v\in X\mid \exists u[(u,v)\in M]\}$.
A matching $M$ is called \emph{perfect} if $M_{(1)}=M_{(2)}=X$.

We call $(v,v)$ a \emph{trivial pair} and we first include all trivial pairs to $M$. We then eliminate the trivial perfect matching $M'=\{(u,u)\mid u\in X\}$ from our consideration by introducing the following restriction.
Given any subset $M' \subseteq M$ and two elements $x,y\in X$, we say that $x$ is \emph{linked to} $y$ in $M'$ if there exists a series $z_1,z_2,\ldots,z_t\in X$ with a certain odd number $t\geq1$ such that $(x,z_1),(z_t,y)\in M'$ and $(z_i,z_{i+1})\in M'$ for any index $i\in[t-1]$.
For any subset $R$ of $X$, we say that $R$ is \emph{uniquely connected to} $X-R$ in $M$ if, for any element $v\in R$, there exist two unique elements $u_1,u_2\in X-R$ such that $(v,u_1),(u_2,v)\in M$.

As the desired variant of $\mathrm{2DM}$, we introduce the following decision problem and study its computational complexity.

\s
{\sc Almost All Pairs 2-Dimensional Matching Problem with Trivial Pairs} ({\sc AP2DM}):
\renewcommand{\labelitemi}{$\circ$}
\begin{itemize}\vs{-1}
  \setlength{\topsep}{-2mm}%
  \setlength{\itemsep}{1mm}%
  \setlength{\parskip}{0cm}%

\item {\sc Instance:} a finite set $X$, a subset $R$ of $X$, and a subset $M \subseteq  X\times X$ including all trivial pairs such that $R$ is uniquely connected to $X-R$ in $M$.

\item {\sc Question:} is it true that, for any distinct pair $v,w\in X$, if either $v\notin R$ or $w\notin R$, then there exists a perfect matching $M_{vw}$ in $M$ for which $v$ is linked to $w$ in $M_{vw}$?
\end{itemize}

For technicality, all entries of $X$ are assumed to be expressed using $O(\log|X|)$ binary symbols. A natural size parameter $m_{set}$ is then defined as  $m_{set}(\pair{X,R,M})=|X|$.

Let $k\geq2$. An instance $(X,R,M)$ to $\mathrm{AP2DM}$ is said to be \emph{$k$-overlapping} if
(i) for any $v\in X$, $|\{u\in X \mid (u,v)\in M\}|\leq k$ and (ii) for any $u\in X$, $|\{v\in X \mid (u,v)\in M\}|\leq k$. When all instances $(X,R,M)$ given to $\mathrm{AP2DM}$ are limited to those that are $k$-overlapping, we call the resulting problem from $\mathrm{AP2DM}$ by $\mathrm{AP2DM}_{k}$.

We intend to show that $\mathrm{AP2DM}_4$ parameterized by $m_{set}$, $(\mathrm{AP2DM}_4,m_{set})$, is inter-reducible to $(\mathrm{2SAT}_3,m_{vbl})$ by short $\dl$-T-reductions.

\begin{theorem}\label{AP2DM-2SAT-short}
$(\mathrm{AP2DM}_{4},m_{set}) \equiv^{\mathrm{sL}}_{T} (\mathrm{2SAT}_3,m_{vbl})$.
\end{theorem}

\begin{yproof}
For the easy of the description of the proof, we use $(3\dstcon,m_{ver})$ instead of $(\mathrm{2SAT}_3,m_{vbl})$ because $(\mathrm{2SAT}_3,m_{vbl})\equiv^{\mathrm{sL}}_{T} (3\dstcon,m_{ver})$ by Lemma \ref{DSTCON-equiv-2SAT}(3).

As the first step, we wish to verify that $(\mathrm{3DSTCON},m_{ver})\leq^{\mathrm{sL}}_{m} (\mathrm{AP2DM}_{4},m_{set})$ although this is a stronger statement than what is actually needed for our claim (since $\leq^{\mathrm{sL}}_{m}$ implies $\leq^{\mathrm{sL}}_{T}$).
Let $(G,s,t)$ be any instance given to $\mathrm{3DSTCON}$ with $G=(V,E)$.  Notice that $G$ has degree at most $3$. To simplify our argument, we slightly modify $G$ so that $G$ has no vertex whose indegree is $3$ or outdegree is $3$.
For convenience, we further assume that $s$ and $t$ are of degree $1$.
Notationally, we write $V^{(-)}$ for $V-\{s,t\}$ and assume that $V^{(-)}$ is of the form $\{v_1,v_2,\ldots,v_n\}$ with $|V^{(-)}|=n$.

Let us construct a target instance $(X,R,M)$ to which we can reduce $(G,s,t)$ by an appropriately chosen short $\dl$-m-reduction. For any index $i\in\{0,1,2\}$, we prepare a new element of the form $[v,i]$ for each $v\in V$ and define $X_i$ to be $\{[v,i]\mid v\in V^{(-)}\}$. The desired universe $X$ is set to be $\{s,t\}\cup X_0\cup X_1\cup X_2$.
As subsets of $X\times X$, we define the following seven sets:  $M_0=\{([v,0],[w,0])\mid u,w\in V^{(-)}, (v,w)\in E\}$, $M_1=\{([v_i,1],[v_{i+1},1]), ([v_{i+1},1],[v_i,1]) \mid i\in[n-1]\}$, $M_2=\{([v_{i+1},2],[v_i,2]), ([v_i,2],[v_{i+1},2]) \mid i\in[n-1]\}$,
$M_3=\{([v,2],[v,0]), ([v,0],[v,1]) \mid v\in V^{(-)}\}$,
$M_4=\{([v_1,1],s), ([v_n,1],s), (t,[v_1,2]), (t,[v_n,2])\}$,
$M_5=\{(s,[u,0]), ([v,0],t)\mid (s,u),(v,t)\in E\}$, and
$M_6=\{(\tilde{u},\tilde{u})\mid \tilde{u}\in X\}$. Finally, $M$ is defined to be the union $\bigcup_{i=0}^{6} M_i$ and $R$ is set to be $\{[v,0]\mid v\in V^{(-)}\}$. Note that $R$ is uniquely connected to $X-R$ because of $M_3$.

To illustrate the aforementioned construction, let us consider a simple example of $G=(V,E)$ with $V=\{v_1,v_2,v_3,v_4,s,t\}$ and $E=\{(s,v_2), (v_3,v_2), (v_2,v_4), (v_4,v_3), (v_3,t)\}$. The universe $X$ is the set $\{s,t,v_i[j]\mid i\in[4],j\in[3]\}$. The constructed $M$ from $G$ is illustrated in Fig.~\ref{fig:AP2DM-graph-figure}.

\begin{figure}[t]
\centering
\includegraphics*[height=2.4cm]{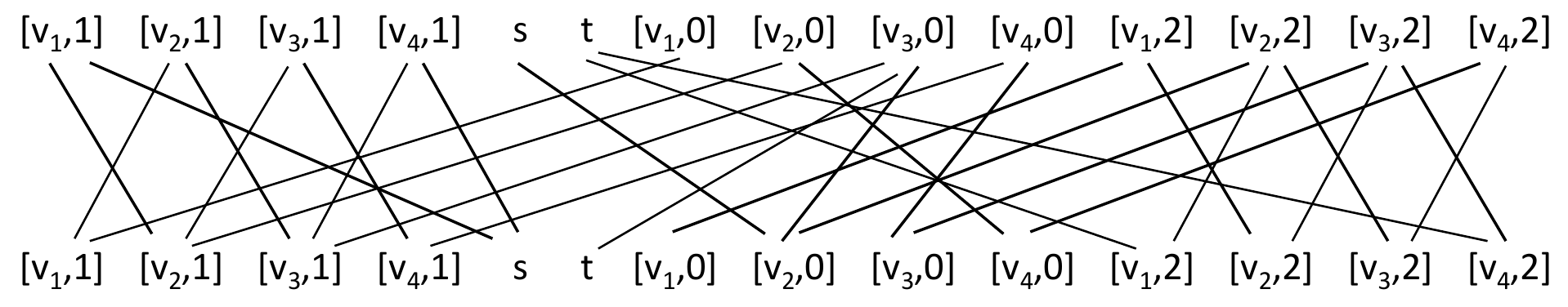}
\caption{The subset $M$ of $X\times X$ with $X=\{s,t,v_i[j]\mid i\in[4],j\in[0,2]_{\integer}\}$ (seen here as a bipartite graph) constructed from $G=(V,E)$ with $V=\{v_1,v_2,v_3,v_4,s,t\}$ and $E=\{(s,v_2), (v_3,v_2), (v_2,v_4), (v_4,v_3), (v_3,t)\}$. Every pair of two adjacent vertices forms a single element in $X$. The edges expressing trivial pairs are all omitted for simplicity. Every vertex has degree at most $4$ (including one omitted edge).}\label{fig:AP2DM-graph-figure}
\end{figure}

In what follows, we claim that there is a simple path from $s$ to $t$ in $G$ iff, for any two distinct elements $\tilde{u},\tilde{v}\in X$, there is a perfect matching, say, $M_{\tilde{u}\tilde{v}}$ for which $\tilde{u}$ is linked to $\tilde{v}$.
To verify this claim, we first assume that there is a simple path  $\gamma_{st}=(w_1,w_2,\ldots,w_k)$ in $G$ with $w_1=s$ and $w_k=t$.
Let
$T=\{([v,0],[w,0])\mid v,w\in\gamma_{st}-\{s,t\}, (v,w)\in E\}$ and $S=\{(s,[w_2,0]),([w_{k-1},0],t)\}$.
We remark that $s$ and $t$ are linked to each other in $M$ because there exists a path $s\leadsto t$ in $G$.
Hereafter,  $\tilde{u}$ and $\tilde{v}$ denote two arbitrary distinct elements in $X$ with either $\tilde{u}\notin R$ or $\tilde{v}\notin R$.

(1) Let us consider the case where $\tilde{u},\tilde{v}\notin \{s,t\}$. In this case, let $\tilde{u}=[v_{i_0},l]$ and $\tilde{v}=[v_{j_0},l']$ for  $l,l'\in[0,2]_{\integer}$ and $i_0,j_0\in[n]$.
It follows that $(l,i_0)\neq (l',j_0)$.
We then define the desired perfect matching $M_{\tilde{u}\tilde{v}}$ as follows, depending on the choice of $\tilde{u}$ and $\tilde{v}$.

(Case 1) Consider the case of $l,l'\in\{1,2\}$.
Let $M'_0=T$, $M'_1=\{([v_i,1],[v_{i+1},1])\mid i\in[n-1]\}$, $M'_2= \{[v_{i+1},2],[v_i,2])\mid i\in[n-1]\}$, $M'_3= \{([v_1,2],[v_1,0]), ([v_1,0],[v_1,1])\}$, $M'_4= \{([v_n,1],s), (t,[v_n,2])\}$, $M'_5= S$, and let $M'_6$ contain $(z,z)$ for all other elements $z$. Finally, we set $M_{\tilde{u}\tilde{v}} =\bigcup_{i=0}^{6} M'_i$.
It then follows by the definition that $M_{\tilde{u}\tilde{v}}$ is a perfect matching. Since $s$ is linked to $t$ in $M_{\tilde{u}\tilde{v}}$, $\tilde{u}$ and $\tilde{v}$ are also linked to each other.

(Case 2) In the case where $l=0$, $l'\in\{1,2\}$, and $v_{i_0}\notin \gamma_{st}$, there are three separate cases (a)--(c) to examine. The symmetric case of Case 2 can be similarly handled and is omitted here.

(a) If $i_0\leq j_0$, then we define $M'_0=T$,
$M'_1=\{([v_i,1],[v_{i+1},1])\mid i\in[i_0,n-1]_{\integer}\}$,
$M'_2=\{ ([v_{i+1},2],[v_i,2])\mid i\in[i_0,n-1]_{\integer}\}$,
$M'_3=\{([v_{i_0},2], [v_{i_0},0]), ([v_{i_0},0],[v_{i_0},1])\}$, $M'_4=\{([v_n,1],s), (t,[v_n,2])\}$, and $M'_5 = S$. We further define $M'_6$ to be composed of $(z,z)$ for all the other elements $z$.
Finally, $M_{\tilde{u}\tilde{v}}$ is set to be the union $\bigcup_{i=0}^{6}M'_i$. Clearly,  $\tilde{u}$ is linked to $\tilde{v}$ in $M_{\tilde{u}\tilde{v}}$.

(b) In the next case of $i_0>j_0$ and $l'=1$, we define $M'_0= T$, $M'_1=\{([v_i,1],[v_{i+1},1])\mid i\in[i_0]\}$,
$M'_2=\{([v_{i+1},2],[v_i,2]) \mid i\in[i_0,n-1]_{\integer}\}$,
$M'_3=\{ ([v_{i_0},2],[v_{i_0},0]), ([v_{i_0},0],[v_{i_0},1]) \}$, $M'_4=\{([v_1,1],s),(t,[v_n,2])\}$, and $M'_5= S$. For all the other elements $z$, we include $(z,z)$ into $M'_6$.  Setting $M_{\tilde{u}\tilde{v}} = \bigcup_{i=0}^{6}M'_i$ makes $\tilde{u}$ be linked to $\tilde{v}$ in it.

(c) In the last case of $i_0>j_0$ and $l'=2$, we define $M'_0= T$, $M'_1=\{([v_i,1],[v_{i+1},1])\mid i\in[i_0,n-1]_{\integer}\}$,
$M'_2=\{([v_{i},2],[v_{i+1},2]) \mid i\in[i_0]\}$,
$M'_3=\{([v_{i_0},2],[v_{i_0},0]), ([v_{i_0},0],[v_{i_0},1]) \}$, $M'_4=\{([v_n,1],s),(t,[v_1,2])\}$, and $M'_5= S$. The set $M'_6$ consists of $(z,z)$ for all the other elements $z$. We then set $M_{\tilde{u}\tilde{v}} = \bigcup_{i=0}^{6}M'_i$.

(Case 3) Consider the case where $l=0$, $l'\in\{1,2\}$, and $v_{i_0}\in \gamma_{st}$. This case is the same as Case 1. In symmetry, the case of $l\in\{1,2\}$, $l'=0$, and $v_{j_0}\in\gamma_{st}$ can be  similarly dealt with.

(Case 4) Consider the case of $l=l'=0$. Assuming that $v_{i_0}\in\gamma_{st}$, we define $M'_0= T$,
$M'_1=\{([v_i,1],[v_{i+1},1])\mid i\in[j_0,n-1]_{\integer}\}$,
$M'_2=\{([v_{i+1},2],[v_i,2]) \mid i\in[j_0,n-1]_{\integer}\}$,
$M'_3=\{ ([v_{j_0},2],[v_{j_0},0]), ([v_{j_0},0],[v_{j_0},1]) \}$, $M'_4=\{([v_n,1],s),(t,[v_n,2])\}$, and $M'_5= S$. As before, we form $M'_6$ by collecting $(z,z)$ for all the other elements $z$. Obviously, $\tilde{u}$ and $\tilde{v}$ are linked in $M_{\tilde{u}\tilde{v}} = \bigcup_{i=0}^{6}M'_i$.

(2) Consider the second case where either $\tilde{u}\in\{s,t\}$ or $\tilde{v}\in\{s,t\}$.
We remark that all the cases discussed in (1) make $s$ (as well as $t$) be linked to any element of the form $[v_{i_0},l]$ and $[v_{j_0},l']$ in the obtained matching. Therefore, we can cope with this case by modifying the construction given (1).

In conclusion, for any $\tilde{u},\tilde{v}\in V$, from (1)--(2), if either $\tilde{u}\notin R$ or $\tilde{v}\notin R$, then there is a perfect matching $M_{\tilde{u}\tilde{v}}$ in which $\tilde{u}$ is linked to $\tilde{v}$.

On the contrary, assume that, for any distinct pair $\tilde{u},\tilde{v}\in X$, there is a perfect matching $M'_{\tilde{u}\tilde{v}}$ in which $\tilde{u}$ is linked to $\tilde{v}$. As a special case, we choose $\tilde{u}=s$ and $\tilde{v}=t$. By the definition of $M$, there is a sequence $(s,[w_1,0],[w_2,0],\ldots,[w_k,0],t)$ such that $(s,[w_1,0]),([w_i,0],[w_{i+1},0]),([w_k,0],t)\in M'_{st}$ for any index $i\in[k-1]$. This implies that $(s,w_1,w_2,\ldots,w_k,t)$ is a path in $G$.


As the second step, it suffices to verify that $(\mathrm{AP2DM}_4,m_{set})\leq^{\mathrm{sL}}_{T} (\mathrm{4DSTCON},m_{ver})$. This is because $(\mathrm{3DSTCON},m_{ver})\equiv^{\mathrm{sL}}_{m} (k\mathrm{DSTCON},m_{ver})$ holds for any $k\geq3$ \cite{Yam17a} by Lemma \ref{DSTCON-equiv-2SAT}(2), and thus the desired reduction of the theorem instantly follows. We start with an arbitrary instance $(X,R,M)$ to $\mathrm{AP2DM}_3$. Remember that $M$ contains all trivial pairs.
We then define a graph $G=(V,E)$ by setting $V=X$ and $E=\{(u,v)\mid u\neq v, (u,v)\in M\}$. Clearly, each vertex in $G$ has degree at most $4$.
Assuming that $M_{uv}$ is a perfect matching, if $u$ is linked to $v$, then $v$ is also linked to $u$.
For any distinct pair $u,v\in X$, if either $u\notin R$ or $v\notin R$, then it follows that there is a perfect matching $M_{uv}$ for which $u$ is linked to $v$ iff there exist one simple path from $u$ to $v$ and another simple path from $v$ to $u$ in $G$. Thus, to check the existence of the desired perfect matching $M_{uv}$, it suffices to make two queries of the forms $(G,u,v)$ and $(G,v,u)$ to $\mathrm{4DSTCON}$ and output YES if the oracle answers affirmatively to the both queries. We then recursively check the existence of $M_{uv}$ for all distinct pairs $u,v\in X$ satisfying either $u\notin R$ or $v\notin R$.
Note that the size $m_{ver}(G,u,v) = m_{ver}(G,v,u) =|V|$ is equal to $m_{set}(X,M)=|X|$. Therefore, we can reduce $(\mathrm{AP2DM}_3,m_{set})$ to $(\mathrm{4DSTCON},m_{ver})$ by short $\dl$-T-reductions.
\end{yproof}

Theorem \ref{AP2DM-2SAT-short} further yields the $\nl$-completeness of $\mathrm{AP2DM}$. Another direct consequence of Theorem \ref{AP2DM-2SAT-short} is the following hardness result for the parameterized decision problem $(\mathrm{AP2DM},m_{set})$.

\begin{corollary}\label{2DM-impossible}
Under LSH, there is no polynomial-time algorithm that solves $(\mathrm{AP2DM},m_{set})$ using $O(m_{set}(x)^{1-\varepsilon})$ space for a certain constant $\varepsilon\in(0,1)$, where $x$ is a symbolic input.
\end{corollary}

\section{A Brief Summary of This Exposition}

Since its first proposal in \cite{Yam17a}, the \emph{linear space hypothesis} (LSH) has been expected to play a key role in showing the computational hardness of numerous combinatorial parameterized-$\nl$ problems.
However, there are few problems that have been proven to be equivalent in computational complexity to $(\mathrm{2SAT}_3,m_{vbl})$. This situation has motivated us to look for natural, practical problems equivalent to $(\mathrm{2SAT}_3,m_{vbl})$.
Along this line of study, the current exposition has introduced three parameterized decision problems $(\mathrm{2CVC}_3,m_{ver})$, $(\mathrm{3XCE}_{2},m_{set})$, and $(\mathrm{AP2DM}_4,m_{set})$, and demonstrated that those problems are all equivalent in power to $(\mathrm{2SAT}_3,m_{vbl})$  by ``short''  log-space reductions.\footnote{We remark that it is unknown that $(\mathrm{AP2DM}_4,m_{set}) \equiv^{\mathrm{sL}}_{m} (\mathrm{2SAT}_3,m_{vbl})$ holds.}
The use of such short reductions is crucial in the equivalence proofs of these parameterized decision problems presented in Sections \ref{sec:vertex-cover}--\ref{sec:2-dimention} because
$\psublin$ is unlikely to be closed under ``standard''  log-space reductions, and short reductions may be more suitable for the discussion on various real-life problems.
Under the assumption of LSH, therefore, all those parameterized decision problems that are equivalent to $(\mathrm{2SAT}_3,m_{vbl})$ by short log-space reductions turn out to be unsolvable in polynomial time using sub-linear space.

In the end, we remind the reader that the question of whether LSH is true still remains open. Nevertheless, we hope to resolve this key  question in the future.



\let\oldbibliography\thebibliography
\renewcommand{\thebibliography}[1]{%
  \oldbibliography{#1}%
  \setlength{\itemsep}{-2pt}%
}
\bibliographystyle{alpha}

\end{document}